\documentclass[12pt]{article}%
\usepackage{amsmath,latexsym}
\usepackage{graphicx}
\usepackage{amsmath}
\usepackage{amsfonts}
\usepackage{amssymb}%
\begin{document}

\title{{Gravitational lensing of wormholes in
   noncommutative geometry}}
   \author{
  Peter K.F. Kuhfittig\\  \footnote{kuhfitti@msoe.edu}
 \small Department of Mathematics, Milwaukee School of
Engineering,\\
\small Milwaukee, Wisconsin 53202-3109, USA}

\date{}
 \maketitle

\begin{abstract}\noindent
It has been shown that a noncommutative-geometry
background may be able to support
traversable wormholes.  This paper discusses
the possible detection of such wormholes in
the outer regions of galactic halos by
means of gravitational lensing.  The procedure
allows a comparison to other models such as
the Navarro-Frenk- White model and $f(R)$
modified gravity and is likely to favor a
model based on noncommutative geometry. \\

\noindent
Keywords: gravitational lensing;  noncommutative geometry
\end{abstract}

\section{Introduction}\label{E:introduction}

Wormholes are handles or tunnels in spacetime
that are able to link widely separated regions of
our Universe or different universes in the
multiverse model.  Such wormholes are described
by the line element \cite{MT88}
\begin{equation}\label{E:line1}
ds^{2}=-e^{2\Phi(r)}dt^{2}+\frac{dr^2}{1-b(r)/r}
dr^{2}+r^{2}(d\theta^{2}+\text{sin}^{2}\theta\,
d\phi^{2}),
\end{equation}
using units in which $c=G=1$.  Here $\Phi=
\Phi(r)$ is the \emph{redshift function}, which must
be everywhere finite to avoid an event horizon,
while $b=b(r)$ is the \emph{shape function}.  For
the shape function, the defining property is
$b(r_{th}) =r_{th}$, where $r=r_{th}$ is the
\emph{throat} of the wormhole.  A critical
requirement is the \emph{flare-out condition}
\cite{MT88}: $b'(r_{th})<1$, while $b(r)<r$
away from the throat.  The flare-out condition
can only be satisfied by violating the null
energy condition.

Another aspect of this paper is noncommutative
geometry: an important outcome of string
theory is the realization that coordinates
may become noncommutative operators on a
$D$-brane \cite{eW96, SW99}.  The commutator
is  $[\textbf{x}^{\mu},\textbf{x}^{\nu}]
=i\theta^{\mu\nu}$, where $\theta^{\mu\nu}$
is an antisymmetric matrix.  Noncommutativity
replaces point-like structures by smeared
objects.  As discussed in Refs. \cite{SSa,SSb},
the smearing effect is accomplished by  using
a Gaussian distribution of minimal length
$\sqrt{\theta}$ instead of the Dirac delta
function \cite{NSS06, pK13}.

In this paper we will assume instead that the
energy density of the static and spherically
symmetric and particle-like gravitational
source has the form
\begin{equation}\label{E:rho1}
  \rho(r)=\frac{M\sqrt{\theta}}
     {\pi^2(r^2+\theta)^2}.
\end{equation}
(See Refs. \cite{LL12, NM08, pKa}.)  Here
the mass $M$ is diffused throughout the
region of linear dimension $\sqrt{\theta}$ due
to the uncertainty.  The noncommutative
geometry is an intrinsic property of spacetime
and does not depend on particular features
such as curvature.

A final topic in this paper is the possible
detection of wormholes in the outer regions
of the galactic halo by means of gravitational
lensing.

\section{Wormhole structure}

That traversable wormholes may exist given a
noncommutative-geometry background is shown
in Ref. \cite{pKb}.  To confirm this
result and to allow a discussion of possible
detection by means of gravitational lensing,
we start with the Einstein field equations:
\begin{equation}\label{E:Einstein1}
  \rho(r)=\frac{b'}{8\pi r^2},
\end{equation}
\begin{equation}\label{E:Einstein2}
   p_r(r)=\frac{1}{8\pi}\left[-\frac{b}{r^3}+
   2\left(1-\frac{b}{r}\right)\frac{\Phi'}{r}
   \right],
\end{equation}
\begin{equation}\label{E:Einstein3}
   p_t(r)=\frac{1}{8\pi}\left(1-\frac{b}{r}\right)
   \left[\Phi''-\frac{b'r-b}{2r(r-b)}\Phi'
   +(\Phi')^2+\frac{\Phi'}{r}-
   \frac{b'r-b}{2r^2(r-b)}\right].
\end{equation}

Next, from Eqs. (\ref{E:rho1}) and
(\ref{E:Einstein1}),
\begin{equation}\label{E:bprime}
   b'(r)=8\pi r^2\frac{M\sqrt{\theta}}
     {\pi^2(r^2+\theta)^2}.
\end{equation}
Since $\theta$ is necessarily small compared
to $M$, it follows immediately that
\[
    0<b'(t_{th})<1,
\]
so that the flare-out condition is satisfied.
Here the small linear dimension $\sqrt{\theta}$
raises a question regarding the scale.  From
Eq. (\ref{E:bprime}),
\begin{multline}\label{E:shape1}
   b(r)=\frac{8M\sqrt{\theta}}{\pi}\int_{r_{th}}
   ^r\frac{(r')^2dr'}{[(r')^2+\theta]^2}+r_{th}\\
   =\frac{4M\sqrt{\theta}}{\pi}\left(\frac{1}
   {\sqrt{\theta}}\text{tan}^{-1}\frac{r}
   {\sqrt{\theta}}-\frac{r}{r^2+\theta}-
   \frac{1}{\sqrt{\theta}}\text{tan}^{-1}
   \frac{r_{th}}{\sqrt{\theta}}
   +\frac{r_{th}}{r_{th}^2+\theta}\right)
   +r_{th}.
\end{multline}
Observe that $b(r_{th})=r_{th}$ (Fig. 1).  Since
\begin{figure}[tbp]
\begin{center}
\includegraphics[width=0.8\textwidth]{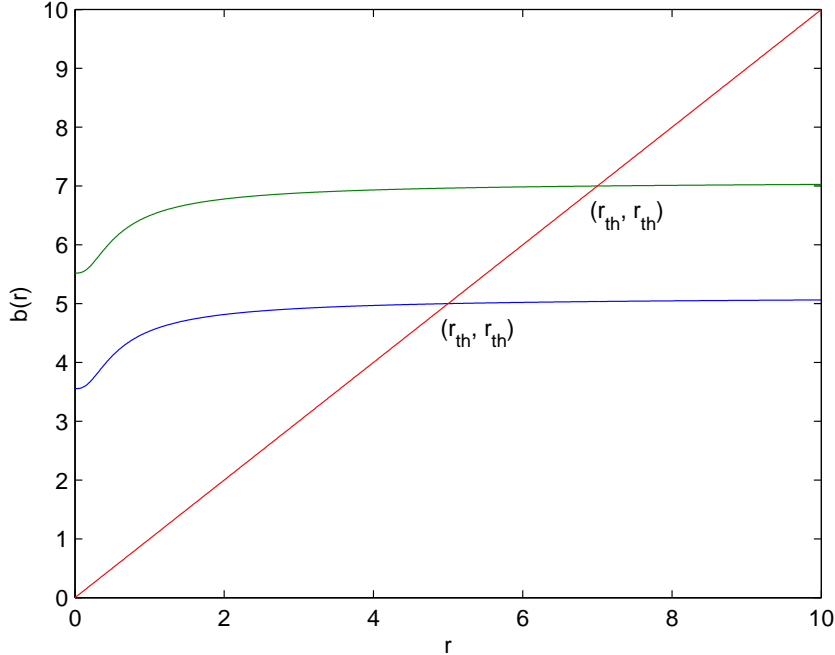}
\end{center}
\caption{The shape function.}
\end{figure}
Eq. (\ref{E:shape1}) is valid for any $r_{th}$,
the wormhole can be macroscopic.  We also
observe that $b(r)<r$ for $r>r_{th}$.

Since we are interested in the detection of
wormholes in the outer regions of the halo, we
need to recall that in this region test particles
move in circular orbits.  So it is assumed that
$e^{2\Phi(r)}$ in line element (\ref{E:line1})
is given by
\begin{equation}
   e^{2\Phi(r)}=\left(\frac{r}{b_0}\right)^l.
\end{equation}
Here $l=2(v^{\phi})^2$, where $v^{\phi}$ is the
tangential velocity and $b_0$ an integration
constant \cite{NVM09}.  According to Ref.
\cite{kN09}, $l=0.000001$.  In a wormhole
setting, we can assume that the center of 
the wormhole also serves as the origin of
$\rho(r)$.  So the line element becomes
\begin{equation}\label{E:line2}
ds^{2}=-\left(\frac{r}{b_0}\right)^ldt^{2}
+\frac{dr^2}{1-b(r)/r}+r^{2}(d\theta^{2}
+\text{sin}^{2}\theta\,d\phi^{2}).
\end{equation}


\section{Gravitational lensing}

Interest in strong gravitational lensing has
increased over time in large part due to a series
of studies by Virbhadra et al. \cite {VNC98, VE00,
CVE01, kV09}.  In particular,  gravitational lensing has been studied from the
strong-field perspective by means of the analytic
method developed by Bozza \cite{vB02} to calculate
the deflection angle.  This method was used in
Refs. \cite{TLa, TLb} to study two special models
by Lemos et al. \cite{jL03}, one of which is the
Ellis wormhole.

To apply these methods to the wormholes in this
paper, we will follow the procedures in Refs.
\cite{TLa, TLb}.  To do so, we will take the
line element to be
\begin{equation}
ds^2=-A(x)dt^2+B(x)dx^2
+C(x)(d\theta^2+\sin^2\theta\,d\phi^2),
\end{equation}
where $x$ is the radial distance in Schwarzschild
units, i.e., $x=r/2M$.  Then
\begin{equation}
    x_0=\frac{r_0}{2M}
\end{equation}
denotes the closest approach to the light ray.
Thus from Eq. (\ref{E:shape1}),
\begin{multline}\label{E:shape2}
   \frac{1}{2M}b(r)=\frac{4M}{\pi (2M)}
   \left(\text{tan}^{-1}\frac{r/2M}
   {\sqrt{\theta}/2M}-
   \frac{(r/2M)(\sqrt{\theta}/2M)}
       {(r^2+\theta)/(2M)^2}\right.\\
    \left.-\text{tan}^{-1}
    \frac{r_{th}/2M}{\sqrt{\theta}/2M}
    +\frac{(r_{th}/2M)(\sqrt{\theta}/2M)}
    {(r_{th}^2+\theta)/(2M)^2}\right)
    +\frac{r_{th}}{2M}.
\end{multline}
In the discussion below, we let $\sqrt{\beta}=
\sqrt{\theta}/2M$.

Now using the lens equation in Ref. \cite{VE00},
it is shown in Refs. \cite{TLa, TLb} that the
deflection angle $\alpha(x_0)$ consists of the
sum of two terms:
\begin{equation}\label{E:angle1}
  \alpha(x_0)=\alpha_e+I(x_0).
\end{equation}
Here
\begin{equation}\label{E:angle2}
   \alpha_e=-2\, \text{ln}\left(\frac{2a}{3}-1\right)
   -0.8056
\end{equation}
is due to the external Schwarzschild metric outside
the wormhole's mouth $r=a$; $I(x_0)$ is the
contribution from the internal metric, first
derived in Ref. \cite{VNC98}:
\begin{equation}\label{E:deflection1}
    I(x_0)=2\int^{\infty}_{x_0}\frac{\sqrt{B(x)}\,dx}
    {\sqrt{C(x)}\sqrt{\frac{C(x)A(x_0)}{C(x_0)A(x)}-1}}.
\end{equation}

From our line element (\ref{E:line2}) and the shape
function (\ref{E:shape2}) [with $\beta=
\theta/(2M)^2$], we obtain
\begin{equation}\label{E:deflection2}
   I(x_0)=\\ \int^a_{x_0}R(x)\,dx,
\end{equation}
where
\begin{multline}
   R(x)=\\ \frac{2}{\sqrt{x^2\left\{1-
   \frac{1}{x}\left[\frac{2}{\pi}
   \left(\text{tan}^{-1}\frac{x}{\sqrt{\beta}}
   -\frac{x\sqrt{\beta}}{x^2+\beta}
   -\text{tan}^{-1}\frac{x_{th}}{\sqrt{\beta}}
   +\frac{x_{th}\sqrt{\beta}}{x_{th}^2+\beta}
   \right)+x_{th}\right]\right\}}
   \sqrt{\frac{x^{2-l}}{x_0^{2-l}}-1}}.
\end{multline}
To see where this integral diverges, we make the
change of variable $y=x/x_0$, so that $x=x_0y$
and $x_{th}=x_0y_{th}$:
\begin{equation}\label{E:deflection3}
    I(x_0)=\int_1^{a/x_0}S(y)\,dy,
\end{equation}
where
\begin{multline}
   S(y)=\\ \frac{2}{\sqrt{(y^{4-l}-y^2)
    \left\{1-\frac{1}{x_0y}\left[\frac{2}{\pi}
    \left(\text{tan}^{-1}\frac{x_0y}{\sqrt{\beta}}
    -\frac{x_0y\sqrt{\beta}}{x_0^2y^2+\beta}
    -\text{tan}^{-1}\frac{x_0y_{th}}{\sqrt{\beta}}
    +\frac{x_0y_{th}\sqrt{\beta}}{x_0^2y_{th}^2+\beta}
    \right)+x_0y_{th}\right]\right\}}}.
\end{multline}
The radicand $F(y)$ in the denominator can be expanded
in a Taylor series around $y=1$.  Letting
\begin{equation}\label{E:form2}
    g(y)=1-\frac{1}{x_0y}\left[\frac{2}{\pi}
    \left(\text{tan}^{-1}\frac{x_0y}{\sqrt{\beta}}
    -\frac{x_0y\sqrt{\beta}}{x_0^2y^2+\beta}
    -\text{tan}^{-1}\frac{x_0y_{th}}{\sqrt{\beta}}
    +\frac{x_0y_{th}\sqrt{\beta}}{x_0^2y_{th}^2+\beta}
    \right)+x_0y_{th}\right],
\end{equation}
we obtain
\begin{multline}
   F(y)=(2-l)g(1)(y-1)\\+
   \left[\frac{1}{2}(5-l)(2-l)g(1)+(2-l)g'(1)\right]
   (y-1)^2\\+
    \text{higher powers}.
\end{multline}
As discussed in Ref. \cite{pK14a}, if $g(1)\neq 0$,
the integral converges due to the leading term
$(y-1)^{1/2}$ resulting from the integration.  If
$g(1)=0$, then the second term leads to
$\text{ln}\,(y-1)$, and the integral diverges.

Next, suppose in Eq. (\ref{E:shape1}), we choose
$r_{th}=r_0$, the nearest approach, for the radius
of the throat.  Since $x_{th}=r_{th}/2M$ and
$y_{th}=x_{th}/x_0$, we now have $y_{th}=1$.
Then
\begin{equation}
g(y)=1-\frac{1}{x_0y}\left[\frac{2}{\pi}
    \left(\text{tan}^{-1}\frac{x_0y}{\sqrt{\beta}}
    -\frac{x_0y\sqrt{\beta}}{x_0^2y^2+\beta}
    -\text{tan}^{-1}\frac{x_0}{\sqrt{\beta}}
    +\frac{x_0\sqrt{\beta}}{x_0^2+\beta}
    \right)+x_0\right].
    \end{equation}
As a result, $g(1)=0$ and the above integral
diverges.  According to Refs. \cite{TLa, TLb},
we have a photon sphere at the throat $r=r_{th}$.
Such a photon sphere can, in principle, be
detected.

\section{Comparison to the NFW model}

It is shown in Ref. \cite{pK14a} that the
Navarro-Frenk-White (NFW) model predicts
a throat size of 0.40 ly.  So if a
wormhole has a throat size that is
significantly different from 0.40 ly, it
cannot be due to the dark matter model,
but it would be consistent with a
noncommutative-geometry background.

An obvious competitor for the dark-matter
model is $f(R)$ modified gravity.  Let us
therefore consider the gravitational field
equations (for the special case
$\Phi'\equiv 0$) proposed by Lobo and
Oliveira \cite{LO12}:
\begin{equation*}
   \rho(r)=F(r)\frac{b'(r)}{r^2},
\end{equation*}
\begin{equation*}
   p_r(r)=-F(r)\frac{b(r)}{r^3}
   +F'(r)\frac{rb'(r)-b(r)}{2r^2}
   -F''(r)\left[1-\frac{b(r)}{r}\right],
\end{equation*}
\begin{equation*}
   p_t(r)=-\frac{F'(r)}{r}\left[1-\frac{b(r)}{r}
   \right]+\frac{F(r)}{2r^3}[b(r)-rb'(r)],
\end{equation*}
where $F=\frac{df}{dR}$.  The curvature scalar is
given by
\begin{equation*}
  R(r)=\frac{2b'(r)}{r^2}.
\end{equation*}
Using these equations, it is shown in Ref.
\cite{pK14b} that to account for dark matter,
it is sufficient for $F$ to be close to
unity and $F'$ to be close to zero.  So if
$\rho(r)$ is taken from the NFW model, the
results are likely to be similar.  This
suggests that a significant departure
from a radius of 0.40 ly can be taken as
evidence for a noncummutative-geometry
background.


\section{Summary}

The first part of this paper confirms that a
noncommutative-geometry background with a static
and spherically symmetric gravitational source
having the form
\begin{equation*}
  \rho(r)=\frac{M\sqrt{\theta}}
     {\pi^2(r^2+\theta)^2}
\end{equation*}
can support traversable wormholes.  Such wormholes
may be located in the outer regions of the
galactic halo.  Using a method for calculating
the deflection angle proposed by Bozza
\cite{vB02}, it is shown that the deflection angle
diverges at the throat, thereby producing a
detectable photon sphere.  By allowing a
comparison to other models, such as the NFW model
and $f(R)$ modified gravity, the procedure is
likely to favor a model based on noncommutative
geometry.

\end{document}